\documentclass[12pt]{spieman}  
\usepackage{amsmath,amsfonts,amssymb}
\usepackage{graphicx}
\usepackage{setspace}
\usepackage{tocloft}
\usepackage{hyperref}

\title{Smartphone tristimulus colorimetry for skin-tone analysis at common pulse oximetry anatomical sites}

\author[a,*]{Joshua A. Burrow}
\author[a,b]{Rutendo Jakachira}
\author[a]{Gannon Lemaster}
\author[a,c,*]{Kimani C. Toussaint, Jr.}
\affil[a]{Brown University, PROBE Lab, School of Engineering, 184 Hope Street, Providence, RI, 02906, USA}
\affil[b]{Brown University, Physics Department, 184 Hope Street, Providence, RI, 02906, USA}
\affil[c]{Brown University Center for Digital Health, 139 Point Street, Providence, Rhode Island, 02912, USA}

\cftpagenumbersoff{figure}
\cftpagenumbersoff{table} 
\begin{document} 
\maketitle

\begin{abstract}\\
\noindent \textbf{Significance:} Smartphones hold great potential in point-of-care settings due to their accessibility and computational capabilities. This is critical as clinicians increasingly seek to quantify skin-tone, a characteristic which has been shown to impact the accuracy of pulse oximetry readings particularly for dark skin tones, and hence, disproportionately affect patient outcomes.\\

\noindent \textbf{Aim:} This study presents a smartphone-based imaging technique for determining individual typology angle (ITA) and compares these results to those obtained using an industry-standard tristimulus colorimeter, particularly for the finger, a common site for pulse oximetry measurements.\\

\noindent \textbf{Approach:} We employ a smartphone-based imaging method to extract ITA values from four volunteers with diverse skin-tones. The study provides recommendations for minimizing errors caused by ambient light scattering, which can affect skin-tone readings.\\

\noindent \textbf{Results:} The smartphone-based ITA (SITA) measurements with camera flash disabled and minimal ambient lighting correlates well with the industry-standard colorimeter without the need for auxiliary adapters and complex calibration. The method presented enables wide-field ITA mapping for skin-tone quantification that is accessible to clinicians.\\

\noindent \textbf{Conclusions:} Our findings demonstrate that smartphone-based imaging provides an effective alternative for assessing skin-tone in clinical settings.
The reduced complexity of the approach presented makes it highly accessible to the clinical community and others interested in carrying out pulse oximetry across a diversity of skin-tones in a manner that standardizes skin-tone assessment. 

\end{abstract}


{\noindent \footnotesize\textbf{*}Joshua A. Burrow, \linkable{joshua$\_$burrow@brown.edu}, Kimani C. Toussaint, Jr., \linkable{kimani$\_$toussaint@brown.edu}}

\begin{spacing}{2}   

\section{Introduction}
\label{sect:intro} 
Variations in skin thickness, temperature and pigment have been identified as confounding factors to pulse oximetry accuracy\cite{Feiner2007,BohndiefSkintoneEffects,ncbi_article,Cabanas2023}.
Overestimation of arterial oxygen saturation in individuals with dark skin-tones, more commonly found in Black, Asian Indian, and Hispanic populations, has led to an increased incidence of undetected hypoxemia, resulting in delayed or missed medical intervention for patients with pulmonary complications\cite{JAMADelayedHopkins}. 
Despite the growing evidence of such biases, pulse oximeters remain widely used due to their non-invasive nature, the opportunity for continuous monitoring and ease of access compared to arterial blood gas (ABG) testing\cite{NEJM_RacialBiasinPulseOx,Hao2024}.
While several alternative approaches are being explored to ameliorate pulse oximetry inaccuracies, particularly in relation to skin tone\cite{Mantri:22,Jakachira:22}, progress has been limited in developing reliable and accurate methods for skin-tone measurement and analysis. 
This challenge is further compounded by the fact that in real-world clinical environments, skin pigmentation is neither systematically measured as part of routine care nor consistently captured in health record systems, making it difficult to apply corrections in practice.
As a result, clinicians rely on misleading proxies such as self-reported race and ethnicity [see examples \cite{valbuena2022racial,BurnettAnestesiologyretro,valbuena2022racial2}]. 
Furthermore, due to the wide range of skin-tones within the Black and Asian Indian communities, using racial profiles for skin phenotyping is unreliable.
Hence, more research is needed to standardize measurement and reporting of skin pigmentation with tools available to all practitioners.
This is especially critical as the U.S. Food and Drug Administration (FDA) aims to specify skin-tone measurement tools and best practices for accurate skin-tone assessment\cite{Vasudevan2024}. 
Historically, skin-tone classification has relied on qualitative scales, such as the von Luschan, Fitzpatrick\cite{fitzpatrick1975soleil} or Martin and Massey\cite{MartinMassey} scales. 
While these methods are extremely practical, they are fundamentally subjective in nature and thus unreliable for clinical purposes\cite{LipnickChallnegeswithSubjective}. 
Moreover, these color stratification scales are either oversampled, optimized for tanning, lack digitized color codes for reproducibility, or often fail to account for variability in the printers used. 
To address these limitations, E. Monk and Google have developed a broader skin-tone stratification system to improve fairness in machine learning applications\cite{Monk_2019}. 
While these phenotypic systems are progressively improving, challenges persist pertaining to subjectivity and failure to capture specific undertones.

Non-perception-based methods, including the use of colorimeters, melanometers, digital cameras and spectrophotometers, provide more objective and reproducible skin-tone measurements, minimizing inherent biases found in perception-based approaches\cite{LY20203}.
However, these devices are expensive, costing \$5,000 to \$20,000 per unit, and vary depending on the manufacturer with respect to device configuration, aperture size and systematic differences in color balancing algorithms. 

\begin{figure}[h!]
\begin{center}
\begin{tabular}{c}
\includegraphics[width=1.00\textwidth]{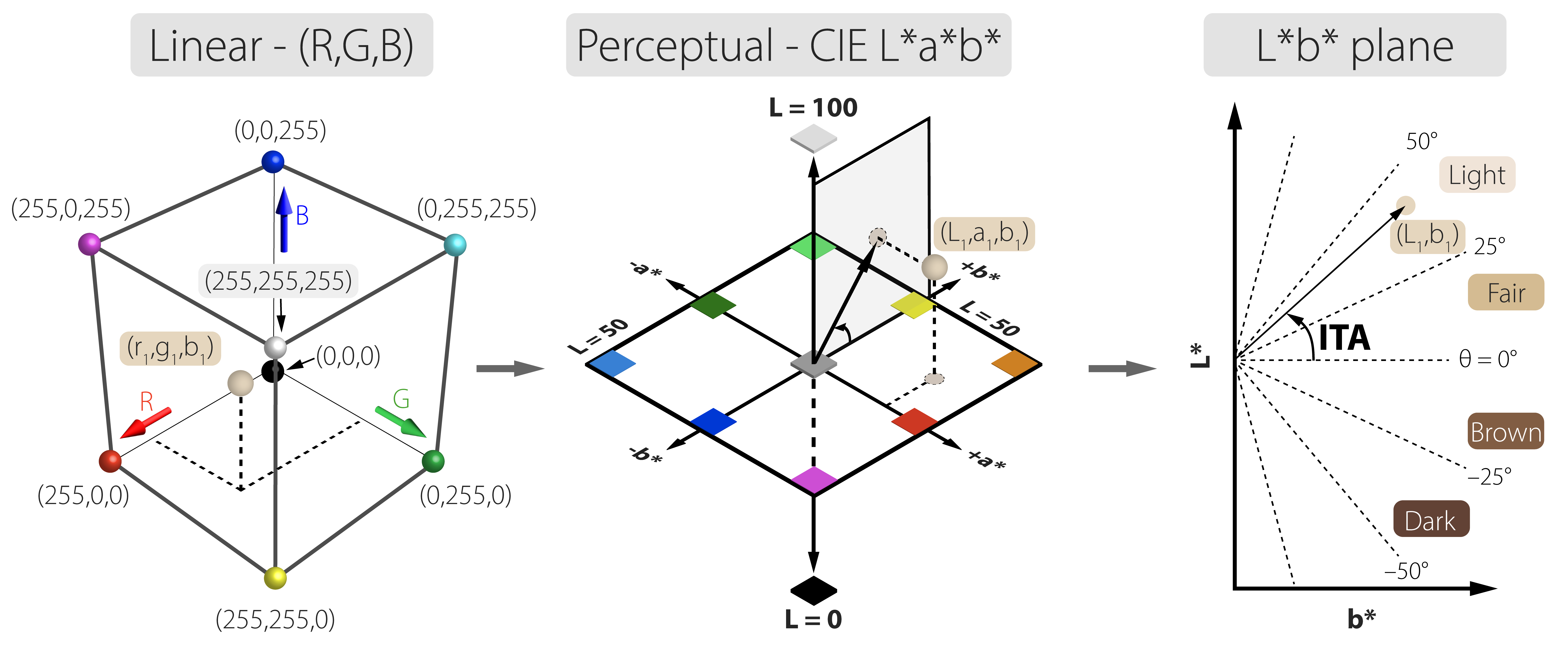}
\end{tabular}
\end{center}
\caption 
{ \label{fig:colorspace}
The three-dimensional linear RGB color space is mapped to the three-dimensional perceptual CIE $L^*a^*b^*$ color space, with an example point marked as $(r_1, g_1, b_1)$ in the RGB space, $(L_1, a_1, b_1)$ in the CIE $L^*a^*b^*$ space, and $(L_1, b_1)$ in the $L^*b^*$ plane for an ITA calculation. The derived ITA from coordinates in the two-dimensional $L^*b^*$ plane and classified using a uniformly distributed phenotypic system, spanning from dark to light. The arbitrary point provides a consistent reference across all three spaces.}
\end{figure}

Nonetheless, these devices operate by measuring diffusely reflected light intensity across multiple visible light wavelengths, typically RGB channels. Figure \ref{fig:colorspace} displays the RGB linear color space which converts color values to the CIE $L^*a^*b^*$ space, a standard in skin-tone analysis due to its perceptual uniformity\cite{DelBinoITAacrossDemograhics}.
The $L^*$ axis quantifies the lightness from 0 (black) to 100 (white), and correlates with the level of pigmentation of the skin. 
The chromaticity coordinate $a^*$ represents the green-red component of a color and correlates with erythema, while the $b^*$ component represents the blue-yellow component and correlates with pigmentation and tanning.
$L^*$ and $b^*$ parameters can be used for constitutive pigment classification according to the dermatological-adopted individual typology angle (ITA) defined as 
\begin{equation}\label{ITAeqn}
\text{ITA}(^{\circ}) = \frac{180^{\circ}}{\pi}\cdot \arctan\bigg(\frac{L^*-50^{\circ}}{b^*}\bigg),
\end{equation}
where $L^*$ and $b^*$ are the luminance and blue-yellow chroma, respectively.
ITA is not skin color, but rather a simplified proxy, since an ITA value can represent many perceived colors with varying radial distances and reddish hues.
Nonetheless, ITA measurements are further categorized into six uniformly distributed skin type groups,\cite{9707510} as opposed to the commonly adopted skin phenotypic scale proposed by Chardon \textit{et al} \cite{chardonPheno}. 

Smartphone-based mobile devices have emerged as powerful tools for assessing personal health information in recent years.\cite{HuntJBO_Smartphoneimaging, Kim:16} 
According to a Pew Research Center survey conducted in 2023, 90\% of U.S. adults own a smartphone.\cite{pew2024}
This widespread adoption has driven substantial research and development efforts toward creating smartphone-integrated healthcare solutions.
Notably, several portable skin diagnostic systems utilizing smartphones have been developed to facilitate the early detection of skin lesions\cite{Kim:16}, as well as pH\cite{PARK2021130359}, urine\cite{urine,smartphoneurine} and glucose monitoring\cite{balbach2021smartphone}. 
Many of these devices rely on RGB color imaging and conventional image processing techniques, yet they face limitations. 
For instance, Cao \textit{et al.} examine the use of smartphone colorimetry to quantify skin-tone under various lighting conditions and exposure settings, highlighting that while smartphones can provide reliable measurements, accuracy varies significantly depending on environmental factors and phone settings, with a focus on optimizing smartphone performance rather than benchmarking against professional devices.\cite{CAOskinsmartphone}
Additionally, Cronin \textit{et al.} demonstrate that smartphone colorimetry can quantify diverse skin-tone tiles at varying camera distances and angles with limited variability in camera distance. 
Their study highlights the importance of controlling for geometric factors, such as distance and angle, to reduce variability in skin-tone measurements.\cite{CroninSmartphoneColorimetrySkinToneTiles}.
Notably, image-based colorimetric methods are convenient and noninvasive, but accuracy is challenging due to differences in lighting conditions located in point-of-care facilities. \cite{ChenQuantifySkinTone2023,Wu1999ImagingCU}


To address these pernicious issues, we present a widely accessible method along with recommendations for skin-tone quantification using a smartphone.
In the proposed approach, the backscattered light from the dorsal and palmar sides of the finger is captured by a cell phone camera across a diverse set of skin-tones under various camera exposure settings and ambient lighting conditions. 
An algorithm is developed to extract an ITA value from an ITA spatial mapping derived from the smartphone-captured RGB photographs and then compared with measurements obtained using an industry-standard colorimeter device. 
Recommended camera settings are provided to facilitate use in a non-laboratory setting, such as clinical environments.

\section{Materials and Methodology}
\label{sect:methods}  

\subsection{Volunteers}
Four healthy volunteers 22 to 35 years of age are included in this investigation.
Volunteers are included such that a diverse distribution of skin pigmentations are assessed.
The skin color of the palmar and dorsal sides of the index finger are collected sequentially using the devices and methods described below.
Since this work focuses on device calibration, it does not meet the federal definition of generalizability and thus does not require IRB approval from the Human Subjects Research Program at Brown University. 


\subsection{DSM-4}
DSM-4 (Cortex Technology) is a handheld diffuse spectroscopic instrument commonly used in clinical applications to perform objective color measurements on skin. 
The anatomical region is illuminated from an angle of 45$^\circ$ by four circularly arranged apparent D65 light sources (see supplemental figure Fig. \ref{fig:supp_figure1} for device spectral response). 
The diffusely reflected light is detected at 0$^\circ$ with respect to the surface normal $\hat{n}$ using a full visible spectrum color sensor (10$^\circ$ standard observer). 
Prior to the measurements performed on each volunteer, the instrument is calibrated by sequential measurements of high-quality white and dark (i.e. zero) standards supplied by the manufacturer. 
Twenty consecutive measurements per anatomical site are captured at a rate of 0.5 Hz. 
Next, the CIE $L^*a^*b^*$ coordinates along with ITA are stored via Bluetooth on the provided software and further analyzed in MATLAB.
The light-tight measurements are not impacted by ambient lighting conditions since the device placed in intimate contact with the skin. 
\begin{figure}[b!]
\begin{center}
\begin{tabular}{c}
\includegraphics[width=1.00\textwidth]{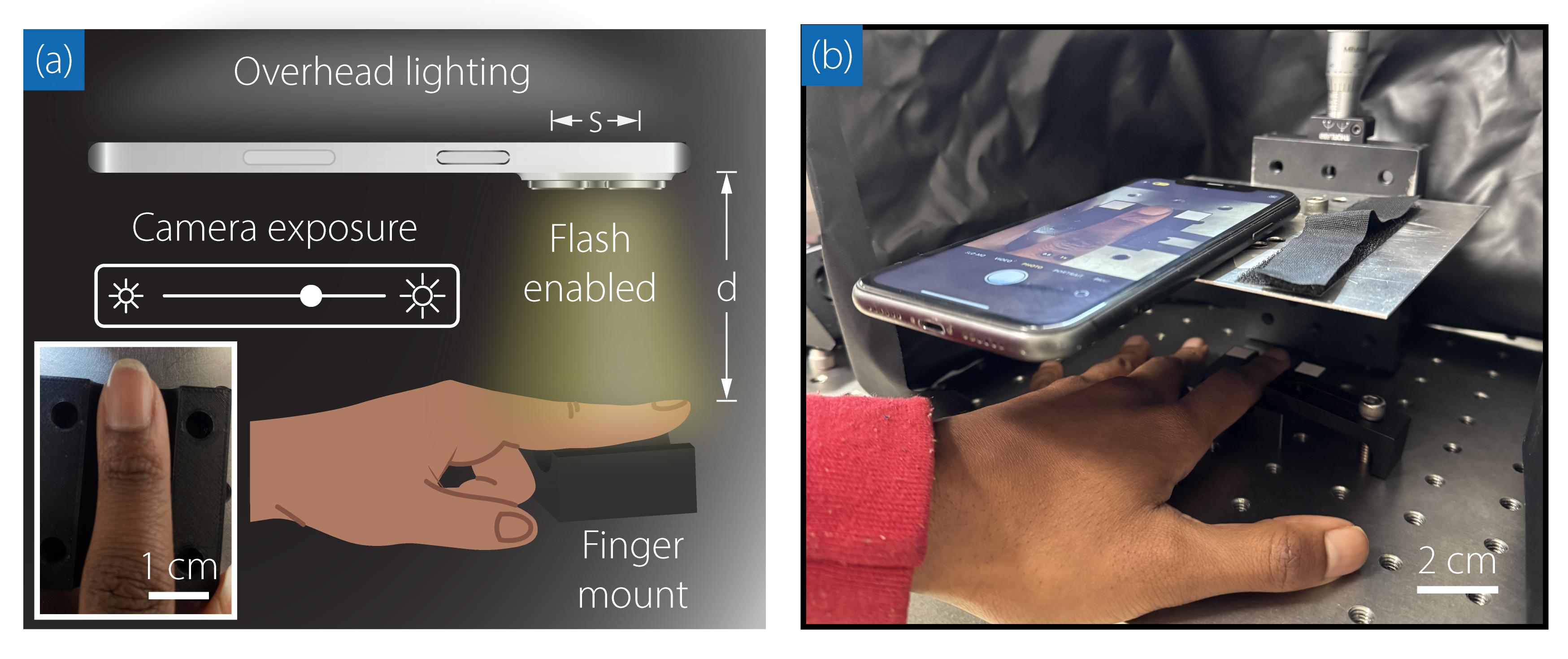}
\end{tabular}
\end{center}
\caption 
{ \label{fig:fig1}
(a) Proposed smartphone colorimeter measured at a common anatomical site for pulse oximetry. The variables $s$ and $d$ are the spacing between the camera flashlight and camera lens, and distance between the smartphone camera and anatomic site, respectively. The inset displays an example photograph captured by the smartphone. (b) Photograph of the Smartphone ITA setup.} 
\end{figure} 

\subsection{Smartphone ITA}
\subsubsection{Smartphone camera}
An iPhone 11 (Apple, Inc.), equipped with a 12-MP camera with an ƒ/1.8 aperture, is employed to capture photographs of skin. 
The camera features allow for substantial light intake and includes a camera flash (see Fig. \ref{fig:supp_figure1} for device spectral response) as shown in Fig. \ref{fig:fig1}(a). The lens and camera flash are separated by a distance $s = 1.5$ cm.
Volunteers place their finger onto a 3D-printed finger mount (see Fig. \ref{fig:supp_figure3}) positioned $d =$ 7 cm away from the smartphone camera. 
The smartphone camera is mounted perpendicular to the target location to capture the image from this fixed distance as displayed in Fig. \ref{fig:fig1}(b).
This simple setup enables color examination of the skin, capturing intricate surface textures, pigmentation variations, and other microstructural features not visible to the naked eye. 
The RGB photographs are captured and further processed using the algorithm described in the next section.

\subsubsection{SITA image analysis algorithm}
\begin{figure}[h!]
\begin{center}
\begin{tabular}{c}
\includegraphics[width=0.5\textwidth]{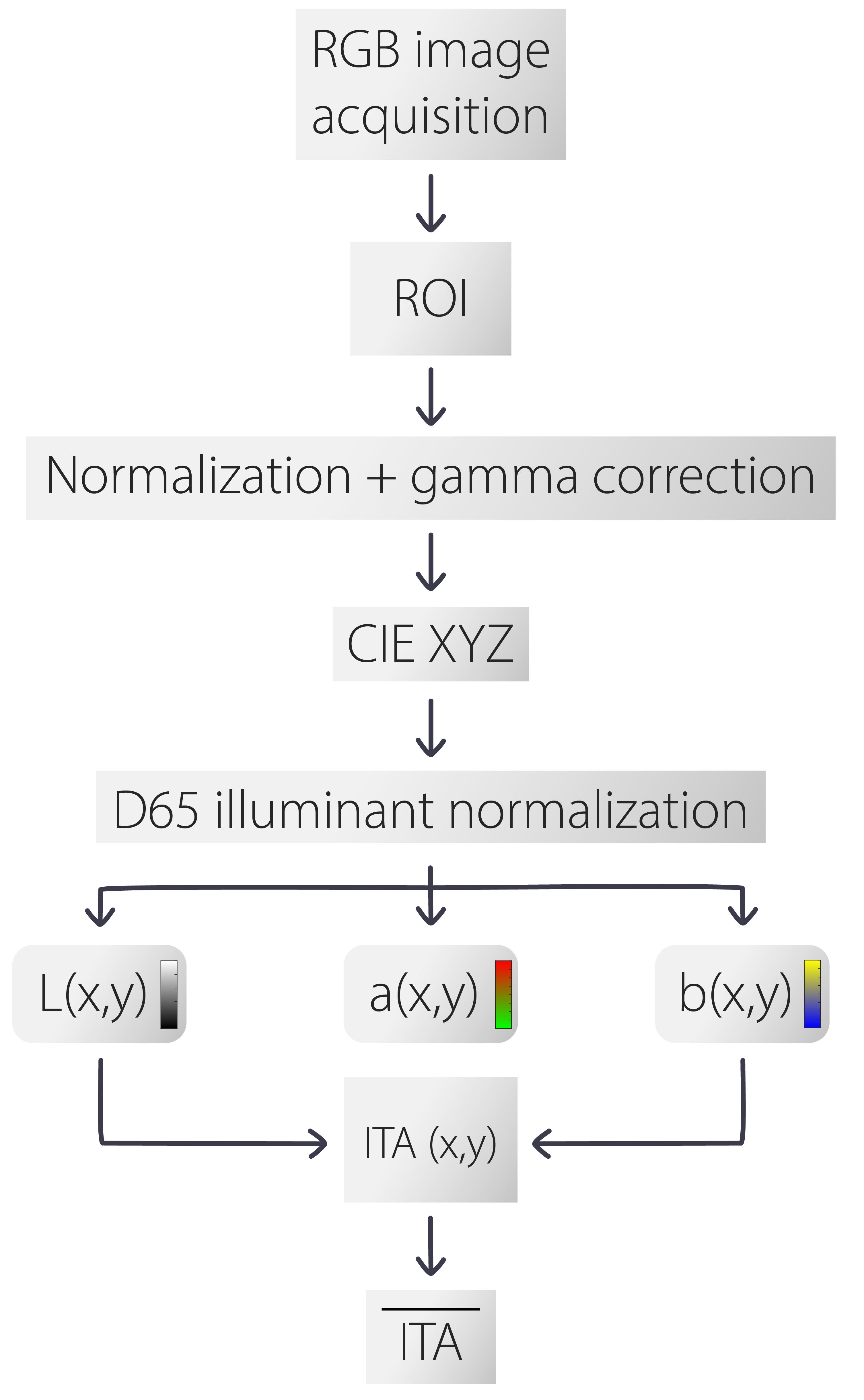}
\end{tabular}
\end{center}
\caption 
{ \label{fig:algoritm}
Flowchart of the proposed image acquisition and SITA algorithm.} 
\end{figure}

Figure \ref{fig:algoritm} displays the workflow of the proposed algorithm. 
The image processing algorithm (herein, referred to as SITA) first selects an $n \times n$ region-of-interest (ROI) from a 3024 $\times$ 3024 pixel 8-bit jpeg image with 72 dpi. 
The transformation of RGB values into the CIE 1976 $L^*a^*b^*$ color space and ultimately a mean ITA value involves several steps. 
This process is essential for converting device-dependent $RGB$ colors into a device-independent color space, which is more perceptually uniform. 
The key steps are detailed below.
First, let $R$, $G$, and $B$ represent the red, green, and blue channels of the input image, respectively, each in the range $[0, 255]$. The RGB values are normalized to the range $[0, 1]$ as follows

\[
    R' = \dfrac{R}{255}, \quad G' = \dfrac{G}{255}, \quad B' = \dfrac{B}{255},
\]
where $R'$, $G'$, and $B'$ are the normalized $RGB$ values. 
The normalized $RGB$ values are then linearized using a gamma correction, as per the $sRGB$ standard \cite{ebner2007gamma,iso22028_2}. The transformation is applied as:
\[
    C(u) = 
    \begin{cases}
        -C(-u), & u < 0 \\
        c\cdot u, & 0 \leq u < 0.0031\\
        a \cdot u^{\gamma} + b, & u \geq 0.0031        
    \end{cases},
\]
where $C'$ is one of the normalized RGB components ($R'$, $G'$, or $B'$), and $C$ is the linearized value for that channel. Here the constants are defined $a = 1.055$, $b = -0.055$, $c = 12.92$ and $\gamma = 1/2.4$.
Next, the linearized $RGB$ values are transformed to the CIE 1931 XYZ \cite{BruceColor} color space using the following matrix transformation
\[
    \begin{pmatrix}
        X \\
        Y \\
        Z
    \end{pmatrix}
    =
    \begin{bmatrix}
        0.4124564 & 0.3575761 & 0.1804375 \\
        0.2126729 & 0.7151522 & 0.0721750 \\
        0.0193339 & 0.1191920 & 0.9503041
    \end{bmatrix}
    \begin{pmatrix}
        R_{\text{lin}} \\
        G_{\text{lin}} \\
        B_{\text{lin}}
    \end{pmatrix},
\]
where $R_{\text{lin}}$, $G_{\text{lin}}$, and $B_{\text{lin}}$ are the linearized RGB values, and $X$, $Y$, and $Z$ are the corresponding tristimulus values in the XYZ color space. The XYZ values are normalized by the D65 illuminant reference white, defined by $X_n$, $Y_n$, and $Z_n$
\[
    X_n = \frac{X}{X_{\text{ref}}}, \quad Y_n = \frac{Y}{Y_{\text{ref}}}, \quad Z_n = \frac{Z}{Z_{\text{ref}}},
\]
where $X_{\text{ref}} = 0.95$, $Y_{\text{ref}} = 1$, and $Z_{\text{ref}} = 1.09$ are the normalized XYZ values. 
The normalized XYZ values are transformed into the LAB color space using the following non-linear function
\[
    f(t) =
    \begin{cases}
        t^{1/3}, & t > \left( \dfrac{6}{29} \right)^3 \\
        \dfrac{1}{3} \left( \dfrac{29}{6} \right)^2 t + \dfrac{4}{29}, & t \leq \left( \dfrac{6}{29} \right)^3
    \end{cases}.
\]
The $L^*$, $a^*$, and $b^*$ values are then calculated as
\[
    L^* = 116 f\left( \frac{Y_n}{Y_{\text{ref}}} \right) - 16,
\]
\[
    a^* = 500 \left[ f\left( \frac{X_n}{X_{\text{ref}}} \right) - f\left( \frac{Y_n}{Y_{\text{ref}}} \right) \right] + \Delta
\]
and \[
    b^* = 200 \left[ f\left( \frac{Y_n}{Y_{\text{ref}}} \right) - f\left( \frac{Z_n}{Z_{\text{ref}}} \right) \right] + \Delta,
\]
where $L^*$ represents the lightness component, $a^*$ and $b^*$ represent the color-opponent dimensions and $\Delta = 128$ for 8-bit images.
The $L^*$ and $b^*$ coordinates are then used to calculate an ITA mapping using Eqn. \ref{ITAeqn}. 
Finally, the mean ITA value represents ITA, calculated by averaging the ITA values accross all pixels in the image. 
This mean ITA, denoted as $\overline{\text{ITA}}$ in Fig. \ref{fig:algoritm}, is used herein as a single representative value for skin-tone.


\section{Results}
\begin{figure}[h!]
\begin{center}
\begin{tabular}{c}
\includegraphics[width=0.85\textwidth]{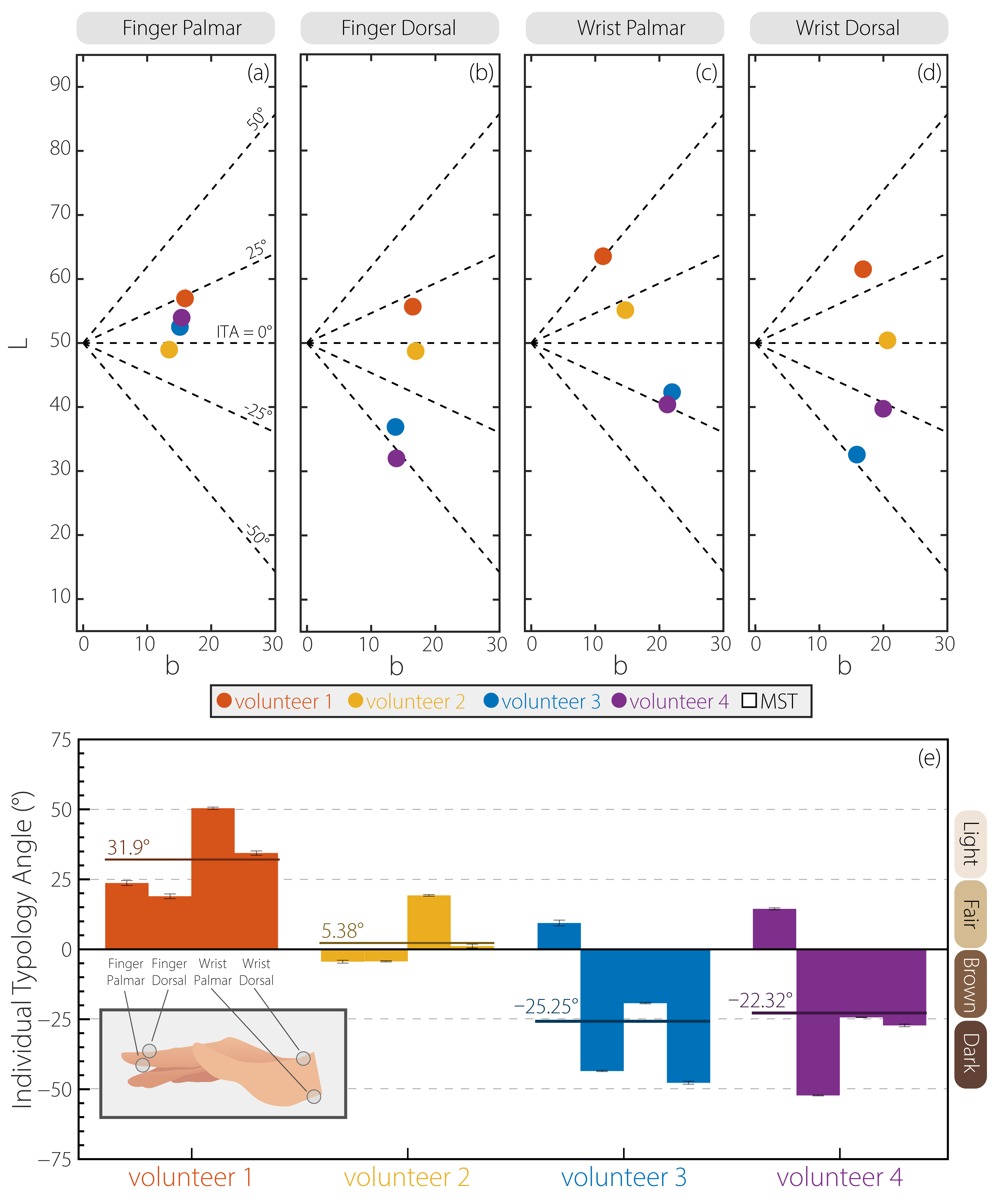}
\end{tabular}
\end{center}
\caption 
{ \label{fig:Lb}
$L$ vs. $b$ for the (a) palmar side of the index finger (with MST overlaid as shaded squares on the palmar finger plot), (b) dorsal side of the index finger, (c) palmar side of the wrist and (d) dorsal side of the wrist. Orange, yellow, blue and purple correspond to volunteers 1, 2, 3, and 4, respectively. Each data point is the $\overline{ITA}$ value from a single individual at a given anatomical site. (e) Individual ITA variation. Dashed horizontal lines represent uniformly distributed ITA cutoffs for skin phenotype.
} 
\end{figure}

Industry-standard ITA values on the palmar and dorsal sides of the finger and wrist are captured with the DSM-4. Figure \ref{fig:Lb}(a)-(d) plots the mean $L^*$ and $b^*$ values for various anatomical sites across the volunteer population. 
The black dashed lines indicate a uniformly distributed phenotypic skin-tone system defined as very light $\in (50^{\circ},90^{\circ}]$, light $\in (25^{\circ},50^{\circ}]$, fair $\in (0^{\circ},25^{\circ}]$, brown $\in (-25^{\circ},0^{\circ}]$, dark $\in (-50^{\circ},-25^{\circ}]$, and very dark $\in [-90^{\circ},-50^{\circ}]$.
Roughly 68.7\% of the mean $(L^{*},b^{*})$ coordinates fall within the commonly adopted banana-shaped CIE-$L^{*}b^{*}$ chromaticity region for skin.
For the palmar side of the finger, the corresponding mean (standard deviation) ITA values for volunteers 1 - 4 are 23.67$^{\circ}$ (0.89$^{\circ}$), 7.93$^{\circ}$ (0.53$^{\circ}$), 7.80$^{\circ}$ (0.34$^{\circ}$), and 14.41$^{\circ}$ (1.04$^{\circ}$), respectively. 
For the dorsal side of the finger, the corresponding mean (standard deviation) ITA values for volunteers 1 - 4 are 18.98$^{\circ}$ (0.83$^{\circ}$), -4.29$^{\circ}$ (0.21$^{\circ}$), -43.50$^{\circ}$ (0.14$^{\circ}$), and -53.21$^{\circ}$ (0.23$^{\circ}$).
A narrower range of ITA angles (-4.41$^{\circ}$ to 23.66$^{\circ}$) are observed among volunteers on the palmar side of the finger spanning 28.07$^{\circ}$, whereas the finger dorsal, palmar and dorsal sides of the wrist span 71.16$^{\circ}$, 75.08$^{\circ}$ and 82.08$^{\circ}$, respectively. 
The ITA values measured at different anatomic sites on the hand within the same participant showed considerable variation [see Fig. \ref{fig:Lb}(e)]. The mean (standard deviation) of all measurements within the same participant are 31.85$^{\circ}$ (13.95$^{\circ}$), 5.38$^{\circ}$ (12.33$^{\circ}$), -25.25$^{\circ}$ (26.30$^{\circ}$), and -22.32$^{\circ}$ (27.52$^{\circ}$). 
In general, a greater degree of variability is observed for darker skin-tones. 


\begin{figure}[h!]
\begin{center}
\begin{tabular}{c}
\includegraphics[width=1.00\textwidth]{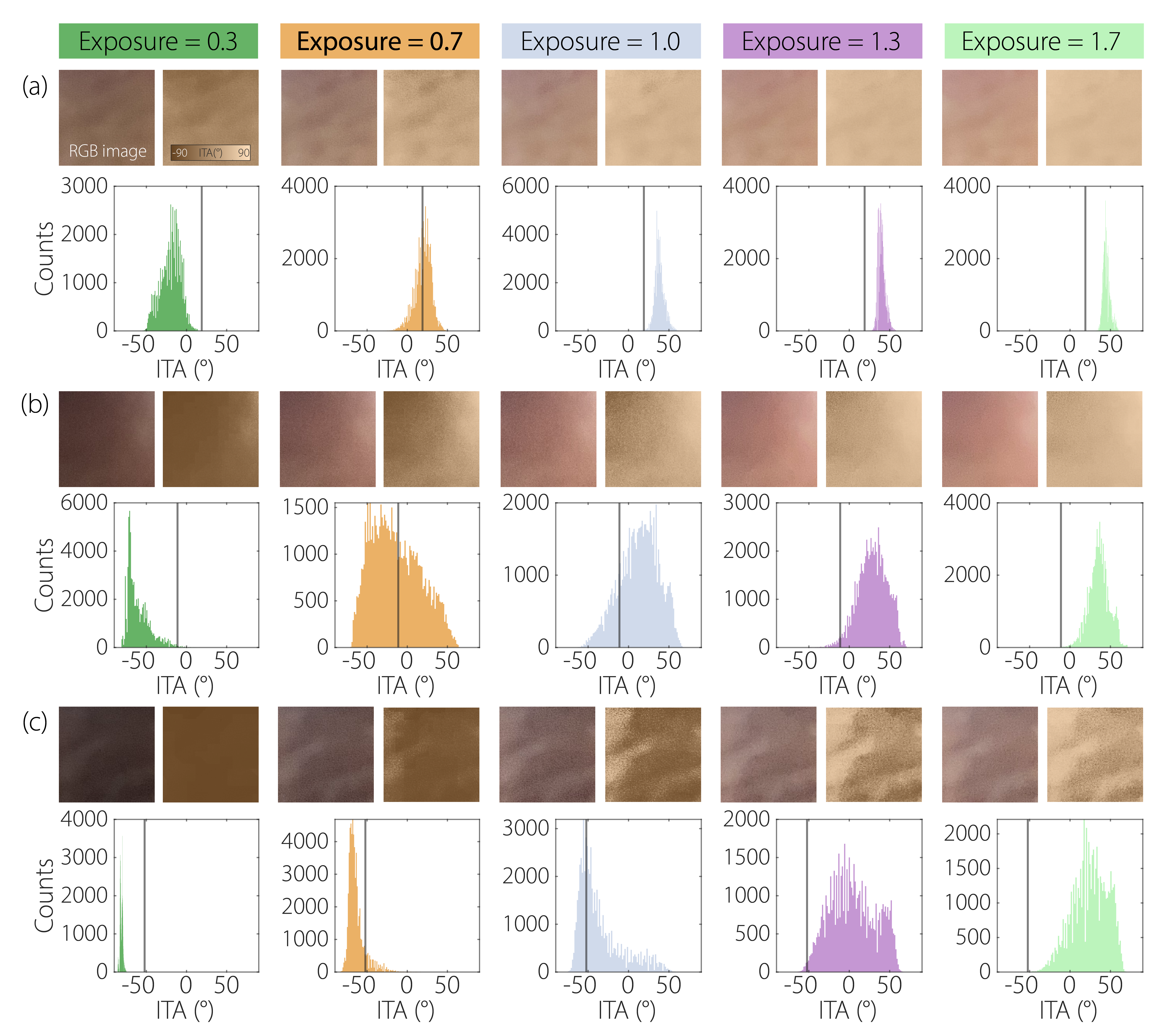}
\end{tabular}
\end{center}
\caption 
{ \label{fig:varyE}
RGB images and ITA colormaps along with the respective ITA histogram for (a) light, (b) fair and (c) dark skin-tones under select camera exposure settings. The industry standard ITA is annotated by a vertical black line in the histograms.
} 
\end{figure} 
Figure \ref{fig:varyE} displays the impact of different exposure settings on the ITA distribution across three distinct skin tones with overhead lights powered off and the camera flash disabled. For each exposure level (0.3, 0.7, 1.0, 1.3 and 1.7), the cropped RGB photographs and calculated ITA spatial mappings are displayed in the top row, with their corresponding ITA histograms displayed below. As exposure increases from 0.3 to 1.7, noticeable shifts occur in both the appearance of the skin tones and the ITA distribution, illustrating how variations in exposure affect the perceived tone and measured characteristics. 
Lower exposure levels (0.3 and 0.7) result in darker skin representations, with ITA histograms displaying negative values, indicating darker skin-tone perceptions. 
Higher exposure levels, particularly at 1.3 and 1.7, yield lighter skin representations and a significant shift in the ITA distribution towards positive values, reflecting a lighter skin tone perception.
This shift is visually apparent in the RGB images and quantitatively evident in the ITA histograms, where the distributions move rightward with increased exposure. 
The ITA histograms further reveal changes in histogram width and peak values across exposure levels. 
In Fig. \ref{fig:varyE}(a), the histogram distributions for the light skin-tone exhibit relatively narrow spreads and higher peaks at lower exposures, while at higher exposures, the ITA distribution broadens with reduced peak counts, indicating increased variability in perceived skin tone. 
Figure \ref{fig:varyE}(b) shows a similar trend for fair skin-tone, with a more pronounced transition from negative to positive ITA values as exposure increases. 
In Fig. \ref{fig:varyE}(c) for a dark skin tone, maintains a narrow ITA distribution at low exposure, but the distribution becomes wider and shifts positively at higher exposures. 
These trends demonstrate that exposure settings can significantly influence ITA-based skin tone measurements, underscoring the need to account for exposure when analyzing or calibrating systems for accurate skin tone representation.

\begin{figure}[h!]
\begin{center}
\begin{tabular}{c}
\includegraphics[width=1.00\textwidth]{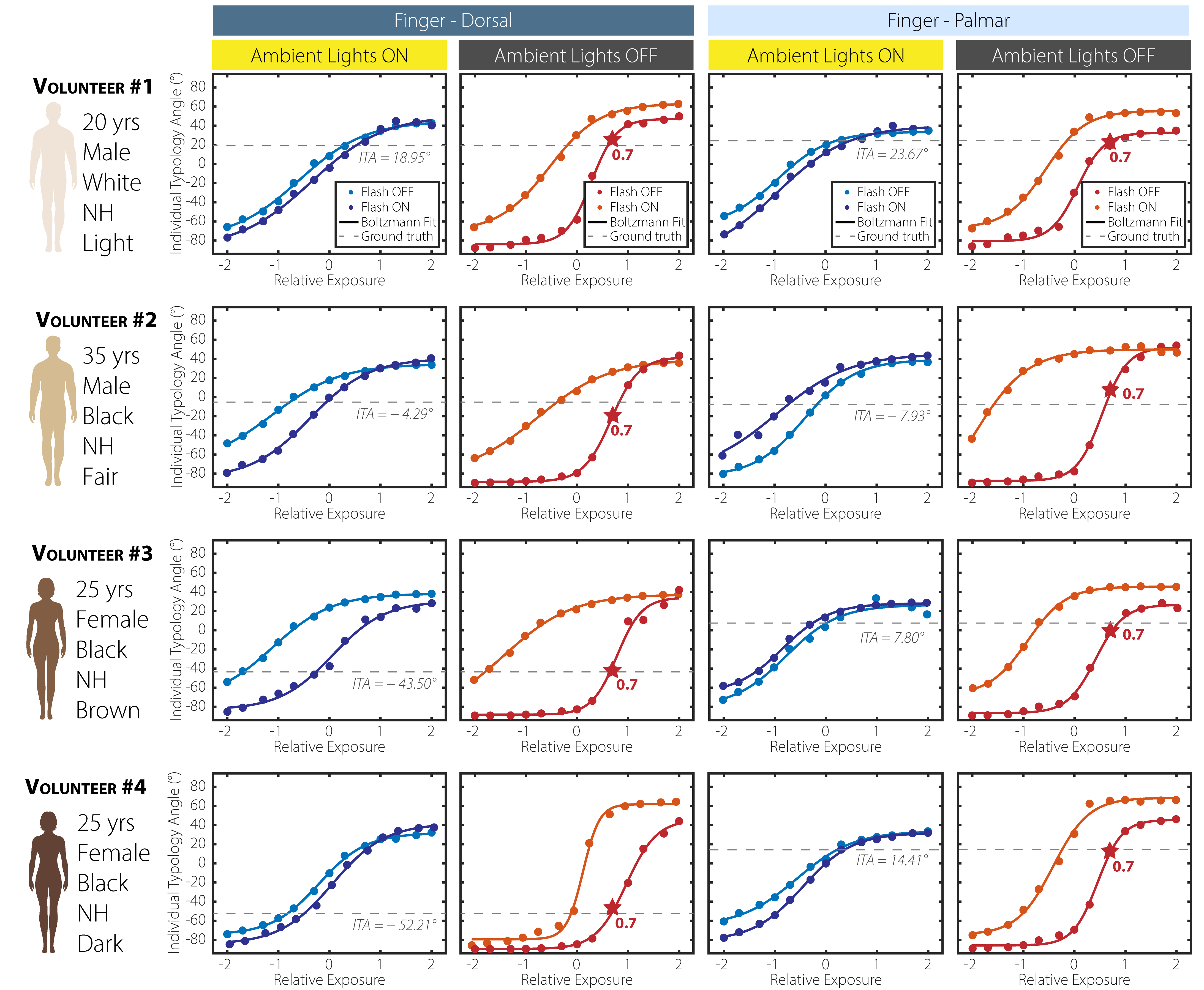}
\end{tabular}
\end{center}
\caption 
{ \label{fig:results}
Volunteer demographics and ITA measurements on the dorsal and palmar side of the finger under varying ambient light and flash conditions. Each row represents a different volunteer with a distinct skin-tone, from light to dark. The blue and orange markers indicate ITA values measured at specific camera exposure settings with flash on and flash off, respectively, while the Boltzmann fitted curves describe the exposure-dependent ITA response. The gray dashed line represents the industry-standard ITA value. The red star on each plot highlights the optimal exposure setting (0.7) for the flash-off and ambient lights-off condition, where ITA values best match the industry standard.}
\end{figure}

To investigate the impact of ambient lighting, sequentially captured image sets are acquired with the room overhead lights on and off. 
Post-image acquisition, the SITA algorithm is then applied to extract a singular ITA value by performing a two-dimensional average of the ITA mapping over the user-selected ROI. The resulting ITA values, expressed as a function of discrete exposure settings, are shown as markers in Fig. \ref{fig:results}.
The first column of Fig. \ref{fig:results} provides the metadata per individual including age, sex, ethnicity and assigned MST score.
A Boltzmann fitting equation is applied to the characteristic curves, defined as $ITA(x) = A_2 + \dfrac{A_1 - A_2}{1 + \exp\left(\frac{x - x_0}{dx}\right)}$, where $A_1$, $A_2$, $x_0$ and $dx$ are fitting parameters corresponding to the minimum (i.e., initial value), maximum (i.e., final value), center and and time constant (i.e., slope), respectively. 
The gray dashed horizontal line represents the mean ITA value measured by an industry-standard colorimeter, providing a gold-standard for comparison across lighting conditions and skin tones.
For measurements performed on the dorsal side of volunteers 1 and 4, a general agreement is observed between the exposure-dependent ITA characteristic curves when the camera flash is enabled (dark blue) and disabled (cyan), with both curves converging near the industry-standard ITA value. 
In contrast, for volunteers 2 and 3, an increase in hysteresis-like behavior is observed when comparing the ITA response under flash-enabled and flash-disabled settings, particularly with ambient lights on. 
This hysteresis-like curve, where ITA values diverge between flash-on and flash-off conditions, suggests increased sensitivity to lighting conditions for medium skin tones compared to light or dark tones. 
Notably, for measurements with the ambient lights off, this hysteresis-like curve is observed across all volunteers, indicating that under controlled lighting, flash usage has a more pronounced effect on ITA responses.
Interestingly, when both flash and ambient lights are off, the ITA responses stabilize across all skin tones, with the optimal exposure setting consistently identified as 0.7, marked by a red star on each plot. 
This consistency is observed in both dorsal and palmar measurements, suggesting that the 0.7 exposure setting yields ITA values closest to the industry standard under these conditions. 
Additionally, the slopes of the linear regime in the Boltzmann fits, represented by $dx$ are relatively comparable across volunteers, with a mean and standard deviation for the fitting parameter 0.23 (0.04). 
This uniformity in slope values indicates a similar sensitivity to exposure changes across skin tones when ambient lighting is minimized, enhancing the robustness of ITA measurements. 
The palmar side of the finger exhibits similar behavior, though industry-standard ITA values are consistently higher than those observed on the dorsal side, potentially due to anatomical differences in pigmentation and reflection properties. 
In scenarios with ambient lights on and the camera flash disabled (i.e., red curves), the ITA characteristic curves exhibit behavior analogous to optical density versus exposure curves, where the initial and final values correspond to under- and over-exposure regimes, respectively. 
Here, the optimal exposure setting remains at 0.7, situated within the linear regime, aligning with prior findings in optical exposure studies \cite{fsu_exposure,Oborska-Kumaszynska11}.

Figure \ref{fig:supp_figure2} further demonstrates the optimal smartphone exposure setting across different skin tones and camera configurations, emphasizing that the "ambient lights off and flash off" setting is more resilient to skin tone variation. 
This condition consistently achieves ITA values that approximate the industry standard, highlighting its potential as a standardized approach for ITA measurements in diverse populations. 
These findings underscore the importance of controlling ambient lighting and exposure settings to achieve reliable ITA measurements across varying skin tones, thereby enhancing the accuracy and reproducibility of smartphone-based skin tone assessments.

\section{Discussion}
In this study, the finger was selected as the anatomical measurement site for skin tone assessment due to its relevance in pulse oximetry, where it is the primary site for clinical measurements. 
Measurements were initially focused on the palmar side of the finger; however, this site showed relatively smaller ITA variability, suggesting limited sensitivity to skin tone differences. 
To evaluate the broader applicability of the SITA technique, we included the dorsal side of the finger, which demonstrated a wider range of ITA values, spanning from -51.21$^\circ$ in volunteer 4 to 18.95$^\circ$ in volunteer 1. 
This increased variability in ITA values on the dorsal side enhances the ability of the technique to capture a broader spectrum of skin tones, highlighting the dorsal side as a more versatile and responsive site for skin tone assessments using ITA.
This is especially important as discrepancies in ITA values can lead to inaccurate skin-tone classifications.

Additionally, our decision to use ITA error, rather than the Euclidean color difference (\( \Delta E \)), was guided by the simplicity and ease of interpreting ITA in clinical settings. 
ITA provides a single angular measure that clinicians can readily use without requiring complex colorimetric calculations, which may not be feasible in clinical environments. 
While \( \Delta E \) offers a comprehensive metric of color difference by combining all three color coordinates, the reliance on the \( L^* \) and \( b^* \) coordinates directly correlates with the most relevant aspects of skin tone (lightness and pigmentation), making it a practical choice for standardized and efficient skin tone measurement.


Based on these results, we recommend specific guidelines for medical practitioners seeking to adopt smartphone-based skin tone assessments in clinical settings such as medical-surgical units, intensive care units, or maternity care rooms. 
First, where possible, control ambient lighting by dimming overhead lights or drawing curtains to reduce natural light fluctuations; bedside lighting may be used if consistent across assessments. 
Practitioners should also disable live photo mode to avoid multi-frame captures that can introduce variability. 
Moreover, disabling auto white balance and selecting manual exposure settings can minimize automatic adjustments that may skew skin tone readings. 
For consistent distance and angle, position the camera approximately 7 cm from the target area and ensure perpendicular alignment to reduce geometric distortion. 
Lastly, saving images in a standard format like jpeg ensures compatibility with analysis software, unlike .heic files, which are less widely supported. These recommendations align with recent guidelines for general medical cellphone photography to ensure high-quality, reproducible images in healthcare settings \cite{Zoltiee067663}.

Despite the promising potential of this approach, several limitations should be noted. This proof-of-concept study was conducted outside of a clinical environment, and measurements were performed on a limited demographic sample of young, healthy adults, which does not fully reflect the anatomical diversity encountered in clinical practice. Furthermore, the ITA calculation relied on 8-bit compressed images, which can introduce quantization errors and may not capture the full range of skin tone variations, especially in cases of subtle pigmentation differences. Despite these limitations, this study lays the groundwork for future evaluations of smartphone-based ITA assessments in point-of-care settings, with the goal of developing a robust, low-cost tool for standardized skin tone measurement.

\section{Conclusion}
In conclusion, our findings support the feasibility of using smartphone cameras for skin tone assessment when controlled settings are applied, specifically in the context of improving pulse oximetry accuracy and addressing skin tone bias in clinical diagnostics.
The optimal exposure setting of 0.7 was identified as a reliable configuration for ITA alignment with industry-standard colorimeter values, particularly in flash-off conditions. 
With additional testing and refinement, this approach has the potential to enhance healthcare equity by providing clinicians with a standardized method for quantifying skin tone in a variety of environments. 
Future studies should expand this work to include a broader range of skin tones, anatomical locations, and lighting conditions commonly encountered in clinical settings, which will be essential for developing a reliable, user-friendly tool for real-world applications.
Finally, in future explorations, apps assisted by artificial intelligence can be designed for SITA which can further refine the skin-tone evaluation process by offering tools for accurate color extraction and analysis, ensuring that the data collected is reliable for clinical or research purposes.

\appendix    

\subsection*{Disclosures}
The authors declare no conflicts of interest.

\subsection* {Code, Data, and Materials Availability} 
The datasets supporting the conclusions of this article will be made available on GitHub (\url{https://github.com/jburrow2/SITA_data}). Additional information may be obtained from the authors upon reasonable request. 

\subsection* {Acknowledgments}
The authors acknowledge partial support from the Brown School of Engineering Hazeltine Award, Brown University Center for Digital Health, Brown University Office of the Vice-President for Research, and the Brown Biomedical Innovations to Impact Award. J.A.B. acknowledges support from the Burroughs Wellcome Fund Postdoctoral Enrichment Program under Grant No. 009248. R.J. acknowledges support from the Google Ph.D. fellowship program in Health Research \& in Algorithmic Fairness.

\bibliography{report}   
\bibliographystyle{spiejour}   

\vspace{2ex}\noindent\textbf{Joshua A. Burrow} received his B.S. degrees in mathematics and physics from Morehouse College in 2012. He earned his M.S. degree and Ph.D. from the Electro-Optics \& Photonics Department at the University of Dayton in 2017 and 2021, respectively. He joined the School of Engineering at Brown University as a Hibbitt postdoctoral fellow and is currently a Burroughs Wellcome Fund fellow in the PROBE Lab. His research interests are in active linear and nonlinear light-matter interactions, nanophotonics, meta-optics, polarization microscopy, and biophotonics. 

\vspace{2ex}\noindent\textbf{Rutendo Jakachira} received her B.S. degree in physics from Drew University in 2019 and is currently working towards her Ph.D. in physics at Brown University working in the PROBE Lab. Her current research interest lie in biomedical opto-electronic devices, signal processing and Monte Carlo modeling of turbid media.

\vspace{2ex}\noindent\textbf{Gannon Lemaster} is an undergraduate student at Brown University, majoring in Electrical Engineering. His research interests focus on optoelectronic fabrication, optics, and photonics, where he is actively exploring advanced technologies that drive innovations in light-based devices.

\vspace{1ex}\noindent\textbf{Kimani C. Toussaint, Jr.} is the Thomas J. Watson, Sr. Professor in the School of Engineering at Brown University. He is the senior associate dean in the School of Engineering and director of the Brown University Center for Digital Health. He directs the laboratory for Photonics Research of Bio/Nano Environments (PROBE), an interdisciplinary research group focusing on developing nonlinear optical imaging techniques for the quantitative assessment of biological tissues and novel methods for harnessing plasmonic nanostructures for light-driven control of matter. 

\listoffigures

\end{spacing}

\newpage

\begin{center}
    \LARGE \textbf{Supplementary Information}
\end{center}


\renewcommand{\thefigure}{S\arabic{figure}}
\setcounter{figure}{0} 

\begin{figure}[h]
\centering
\includegraphics[width=1.0\textwidth]{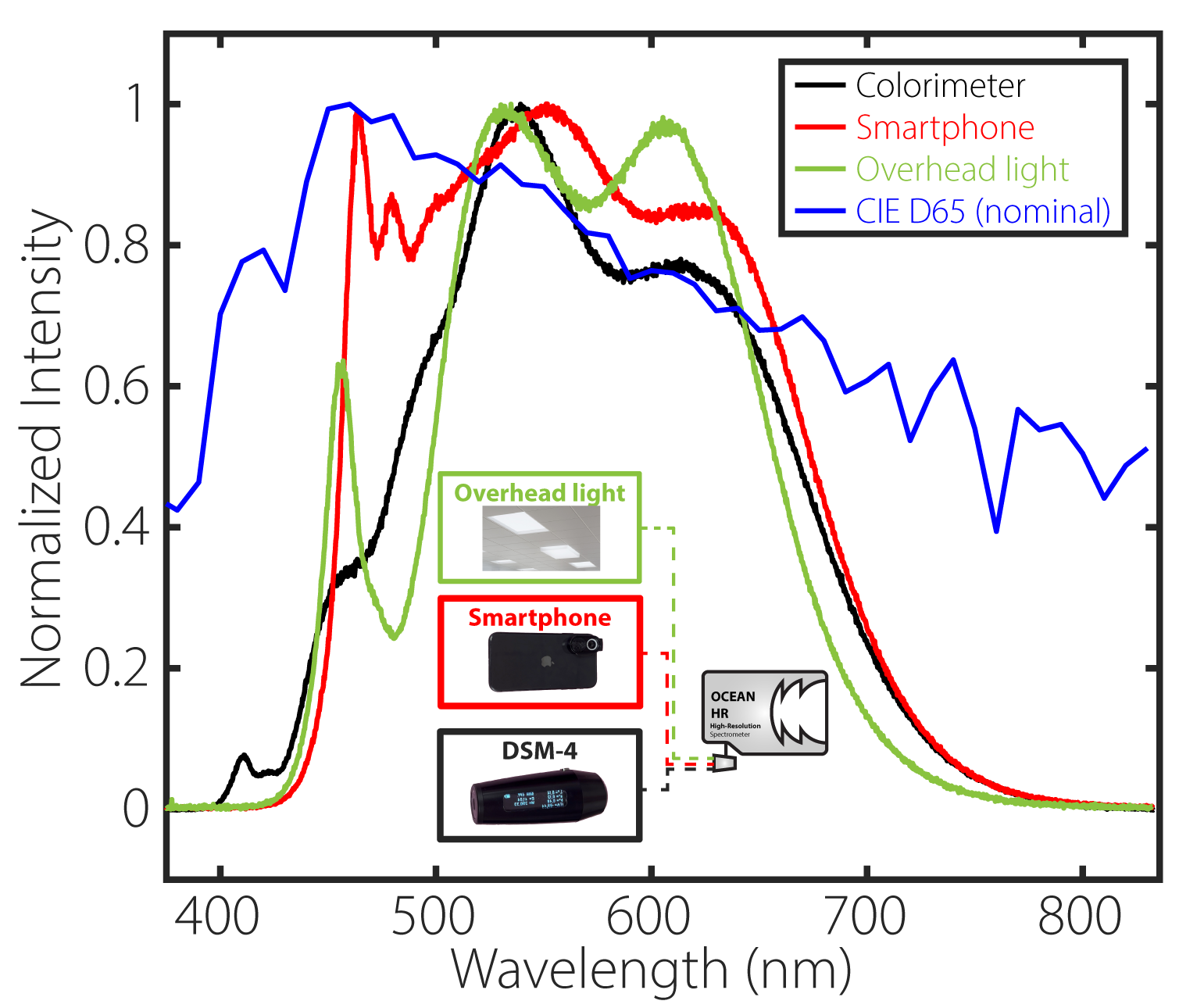}
\caption{Spectral intensity comparison between smartphone (red line) and DSM-4 spectrometer (black line) measurements across the visible spectrum (375 – 825 nm). The smartphone data closely follows the DSM-4 spectrometer's profile, validating its use for colorimetric analysis. Both light sources and the overhead light (green line) do not relatively match with the nominal CIE D65 illuminant spectrum (blue line). Insets show an overhead light (top), the smartphone setup with an attached lens (middle) and DSM-4 spectrometer device (bottom) used for reference measurements.}
\label{fig:supp_figure1}
\end{figure}

\begin{figure}[h]
\centering
\includegraphics[width=0.8\textwidth]{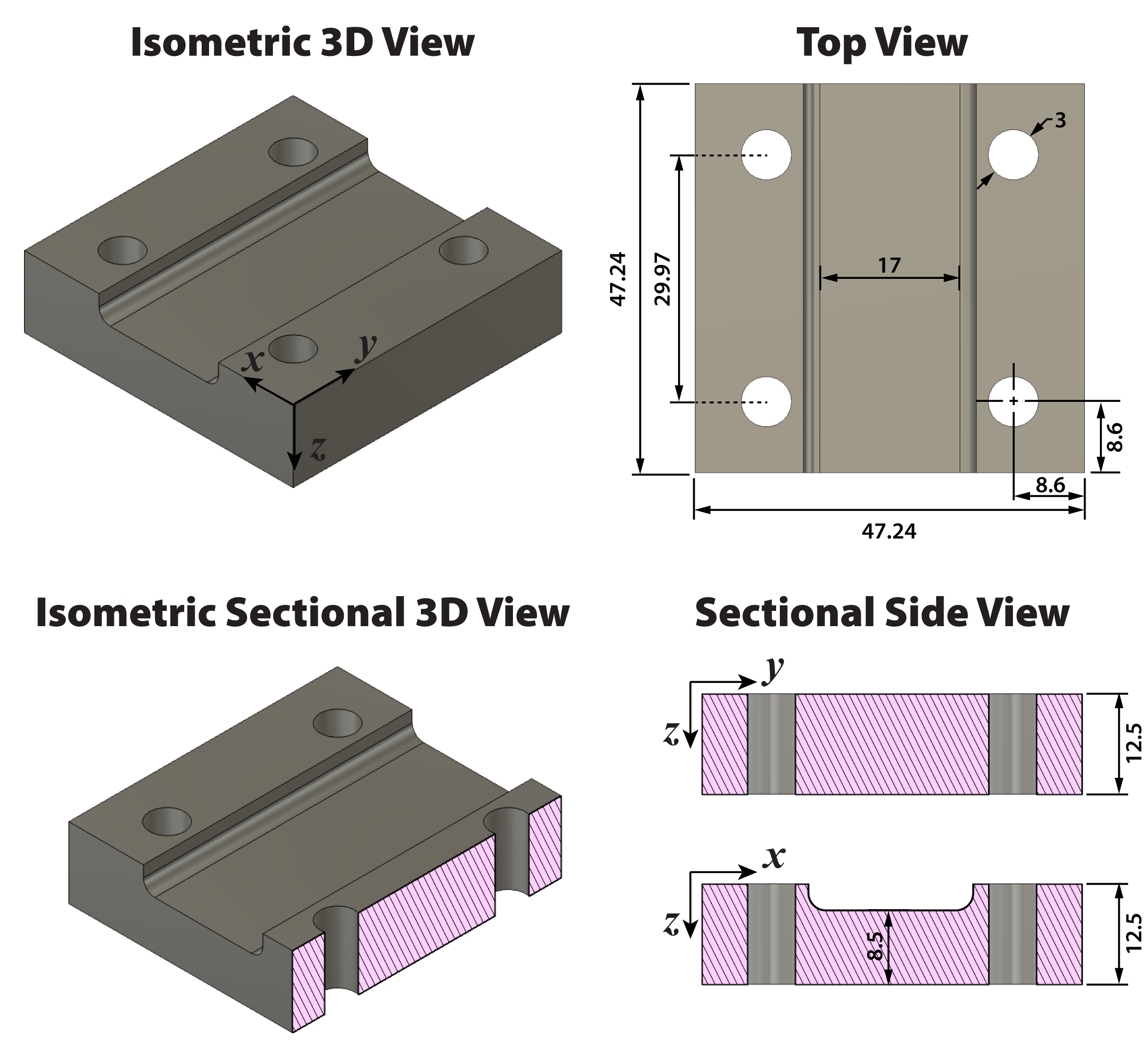}
\caption{Three-dimensional and select orthographic views of the 3D printed finger mount. All measurements in millimeters.}
\label{fig:supp_figure3}
\end{figure}

\begin{figure}[h]
\centering
\includegraphics[width=1.0\textwidth]{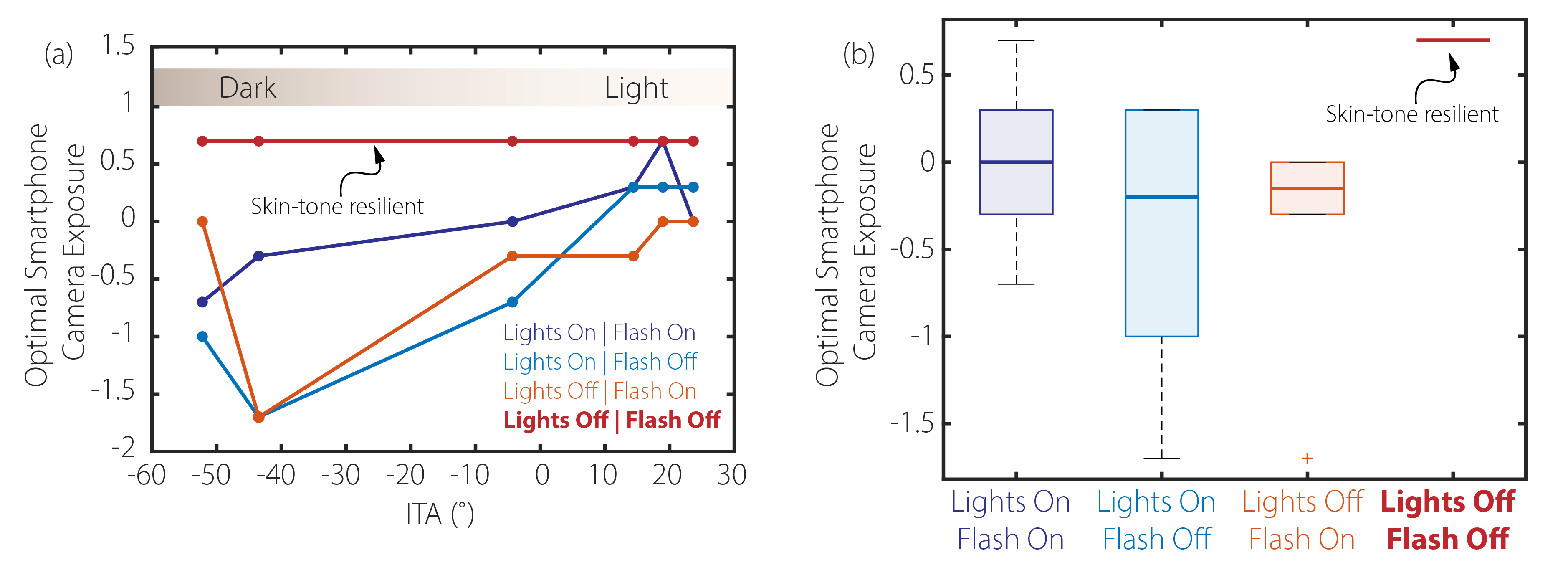}
\caption{Optimal smartphone camera exposure settings for skin tone assessment under different lighting and flash conditions. (a) Optimal exposure settings as a function of ITA for each condition, with the Lights Off | Flash Off configuration (red line) remaining stable across ITA values. (b) Box plot showing the distribution of optimal exposure settings for each condition.}
\label{fig:supp_figure2}
\end{figure}

\end{document}